\documentclass[12pt]{iopart}
\usepackage{graphicx}
\usepackage{eqnarray}
\usepackage{iopams}
\usepackage{cite}

\begin{document}
\title{Electric field control of the skyrmion lattice in Cu$_{2}$OSeO$_{3}$}
\author{J S White$^{1,2}$, I Levati\'{c}$^3$, A A Omrani$^1$, N Egetenmeyer$^2$,\\ K Pr\v{s}a$^1$, I \v{Z}ivkovi\'{c}$^3$, J L Gavilano$^2$, J Kohlbrecher$^2$,\\ M Bartkowiak$^4$, H Berger$^5$, and H M R{\o}nnow$^1$}
\address{$^1$ Laboratory for Quantum Magnetism, Ecole Polytechnique F\'{e}d\'{e}rale de Lausanne (EPFL), CH-1015 Lausanne, Switzerland.}
\address{$^2$ Laboratory for Neutron Scattering, Paul Scherrer Institut, CH-5232 Villigen, Switzerland.}
\address{$^3$ Institute of Physics, Bijeni\v{c}ka 46, HR-10000 Zagreb, Croatia.}
\address{$^4$ Laboratory for Developments and Methods, Paul Scherrer Institut, CH-5232 Villigen, Switzerland.}
\address{$^5$ Institute of Condensed Matter Physics, Ecole Polytechnique F\'{e}d\'{e}rale de Lausanne (EPFL), CH-1015 Lausanne, Switzerland.
\ead{jonathan.white@epfl.ch}}
\begin{abstract}
\\Small-angle neutron scattering has been employed to study the influence of applied electric (E-) fields on the skyrmion lattice in the chiral lattice magnetoelectric Cu$_{2}$OSeO$_{3}$. In an experimental geometry with the E-field parallel to the [111] axis, and the magnetic field parallel to the [1\={1}0] axis, we demonstrate that the effect of applying an E-field is to controllably rotate the skyrmion lattice around the magnetic field axis. Our results are an important first demonstration for a microscopic coupling between applied E-fields and the skyrmions in an insulator, and show that the general emergent properties of skyrmions may be tailored according to the properties of the host system.
\end{abstract}
\section{Introduction}
In recent years there has been a surge of interest in the rich electromagnetic phenomena displayed by topological spin textures. Of special focus are the properties of the skyrmion spin texture, which is manifested as a hexagonal lattice of topologically protected spin `whirls' discovered to occupy a small portion of the magnetic field and temperature phase diagram of the binary (or doped binary) B20 metals MnSi~\cite{Muh09}, FeGe~\cite{Yu11} and Fe$_{1-x}$Co$_{x}$Si~\cite{Mun10,Yu10}. All of these materials display a chiral cubic spacegroup symmetry $P$2$_{1}$3 (No. 198) which allows the Dzyaloshinskii-Moriya interaction to compete with a predominantly ferromagnetic exchange. Below $T_{\rm N}$, this competition results in the stabilization of long wavelength helimagnetic order in zero magnetic field. The skyrmion lattice (SkL) phase is stabilised out of the zero-field state by both a small magnetic field and the gaussian fluctuations close to $T_{\rm N}$~\cite{Muh09}. The magnetic order underlying the SkL is a triple-\textbf{q} state, where each wavevector describes a helix propagating in the plane orthogonal to the applied magnetic field, and at relative angles of 2$\pi$/3. A hallmark feature of the SkL is that it may be stabilised for \emph{any} direction of the magnetic field with respect to the crystal, and is always found to occupy a similar portion of the magnetic phase diagram. This observation was made first on MnSi using the \emph{bulk} probe of small-angle neutron scattering (SANS)~\cite{Muh09}. Furthermore, skyrmion motion in MnSi was demonstrated in the presence of both an applied electric current and a small temperature gradient across the sample~\cite{Jon10}. For a current density above a critical threshold, the forces exerted by the conduction electrons on the skyrmions are sufficient to overcome pinning and cause their coherent motion~\cite{Jon10}. This process was recently clarified to give rise to emergent electrodynamics~\cite{Sch12}.



An important open question was whether skyrmion spin textures remain to be found only within the family of binary B20 metals. Recently, observations by Lorentz transmission electron microscopy~\cite{Sek12}, and further by small-angle neutron scattering~\cite{Ada12,Sek12c}, revealed that a SkL phase exists in the insulating compound Cu$_{2}$OSeO$_{3}$. Since the observed magnetic phase diagram of Cu$_{2}$OSeO$_{3}$ is very similar to that of the B20 alloys, and it furthermore shares the chiral cubic $P$2$_{1}$3 space group symmetry, Cu$_{2}$OSeO$_{3}$ represents a thus far unique example of an insulating analogue for the more well-studied conducting B20 metals. Moreover, Cu$_{2}$OSeO$_{3}$ is well-known to exhibit a clear magnetoelectric (ME) coupling~\cite{Bos08,Mai11,Bel10,Bel12,Ada12,Ziv12,Mai12} which is established by Seki and co-workers~\cite{Sek12,Sek12b} to be active within the SkL phase for certain directions of the magnetic field. Therefore, the next important question is whether the skyrmions in ME insulators may be manipulated directly via the application of electric fields, in an analogous manner to that displayed by applying electric currents to the SkLs in B20 alloys~\cite{Jon10,Sch12}. The answer to this question is important with regard to possible applications since, in principle, skyrmion manipulation in insulators can be done more efficiently than in metals, and without resistive energy losses.

In this Fast Track Communication, we report a small-angle neutron scattering (SANS) study of the microscopic magnetic order underlying the SkL phase in a \emph{bulk} single crystal sample of Cu$_{2}$OSeO$_{3}$. With the sample under simultaneous magnetic and electric fields, we provide the first demonstration of skyrmion manipulation in an insulating compound. The physical picture behind the coupling between electric field and the skyrmions is discussed in light of recent calculations~\cite{Sek12b,Yan12}.


\section{Experimental}
For our experiments a single crystal of Cu$_{2}$OSeO$_{3}$ was grown using a chemical vapour transport method~\cite{Bel10}. The sample had a mass of 26(1)~mg, and volume of 4.5~mm x 2.5~mm x 0.9~mm with the thin axis parallel to the cubic [111] direction. By means of x-ray Laue the sample was oriented so that the horizontal plane for the neutron experiments was defined by both the [111] and [1\={1}0] directions. The crystal was then mounted onto a purpose-built sample stick designed for the application of dc electric fields (E-fields). Electrodes were created directly on the flat faces of the sample using silver paint. The sample stick was loaded into the variable temperature insert of a dedicated horizontal field SANS cryomagnet, which is able to apply magnetic fields of either polarity. The cryomagnet was installed onto the SANS beam line so that the magnetic field was approximately parallel to the neutron beam. The crystal orientation determined by x-ray Laue was aligned with respect to both the magnetic field direction and neutron beam using a bespoke laser system that is accurate to $<$0.1$^{\circ}$.

The SANS measurements were performed using the SANS-I instrument at the Swiss Spallation Neutron Source SINQ, Paul Scherrer Institut, Switzerland. Neutrons of wavelengths $\lambda_{n}=8$~\AA~were selected with a wavelength spread $\Delta\lambda_{n}/\lambda_{n} = 10\%$, and were collimated over distances 15~m-18~m before the sample; the diffracted neutrons were collected by a position-sensitive multidetector placed 15~m-18~m after the sample. The SANS diffraction patterns were measured by rotating the sample and cryomagnet together through an angular range that moved the diffraction spots through the Bragg condition at the detector. All of our measurements in the magnetically-ordered phase were carried out at $T=57$~K since, according to the reported magnetic phase diagrams of Cu$_{2}$OSeO$_{3}$~\cite{Sek12,Ada12}, both the zero field and SkL phases are accessible at this temperature by simply changing the applied magnetic field. Background measurements were carried out in the paramagnetic state at $T=70$~K, and subtracted from the 57~K measurements to leave only the signal due to the magnetism.

For all of our measurements, the E-field remained parallel to the [111] direction, and the magnetic field direction approximately parallel to the neutron beam. A motorised rotation of the sample stick about the vertical axis permitted the study of different magnetic field directions with respect to the crystal in the plane defined by [111] and [1\={1}0]. As such we explored two geometries in our SANS experiments; i) $E\parallel \mu_{0}H \parallel$ [111] and ii) $E\parallel$[111] with $\mu_{0}H \parallel$[1\={1}0].

\section{Results}
\begin{figure}
\centering
  \includegraphics[width=7cm]{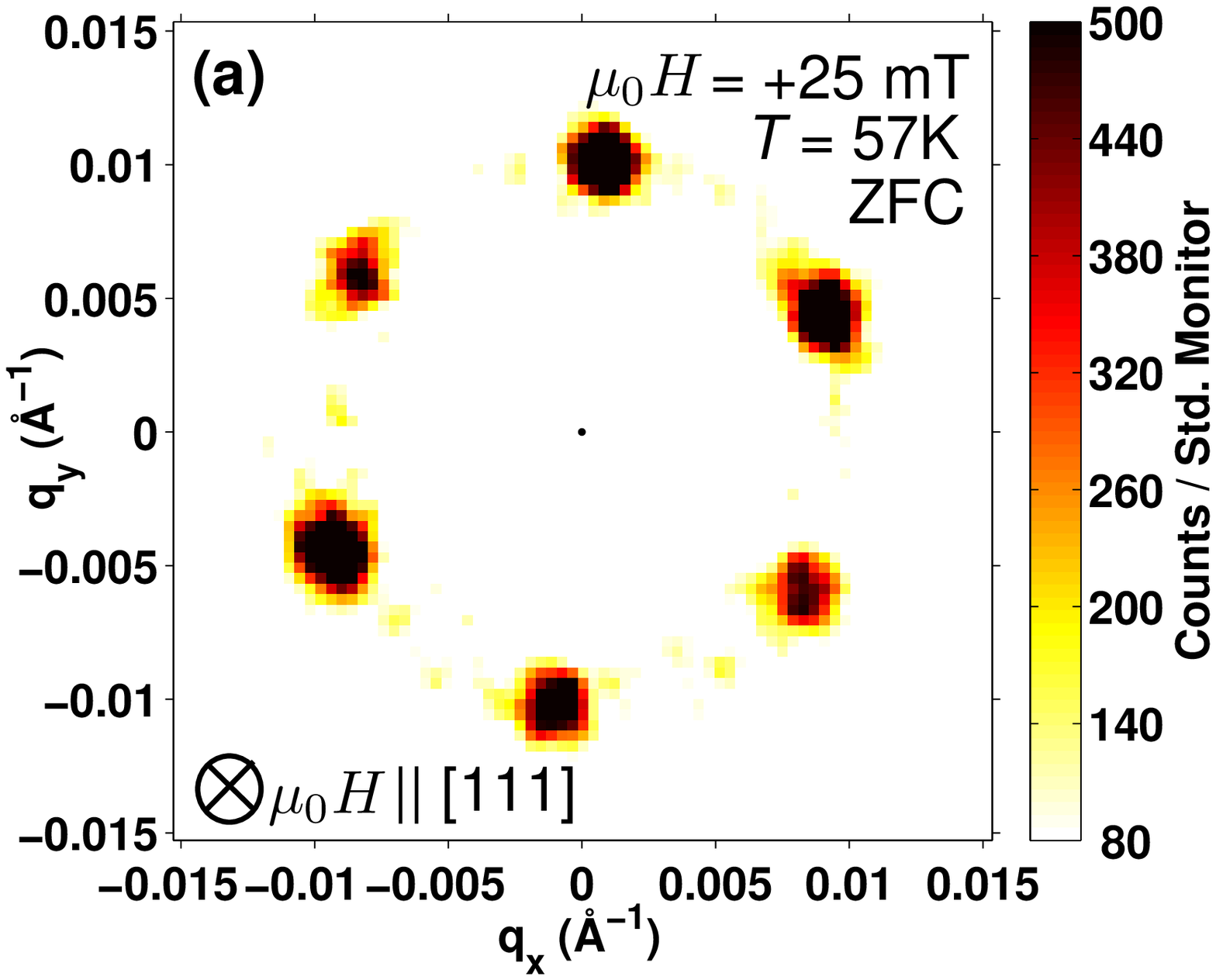}
  \includegraphics[width=7cm]{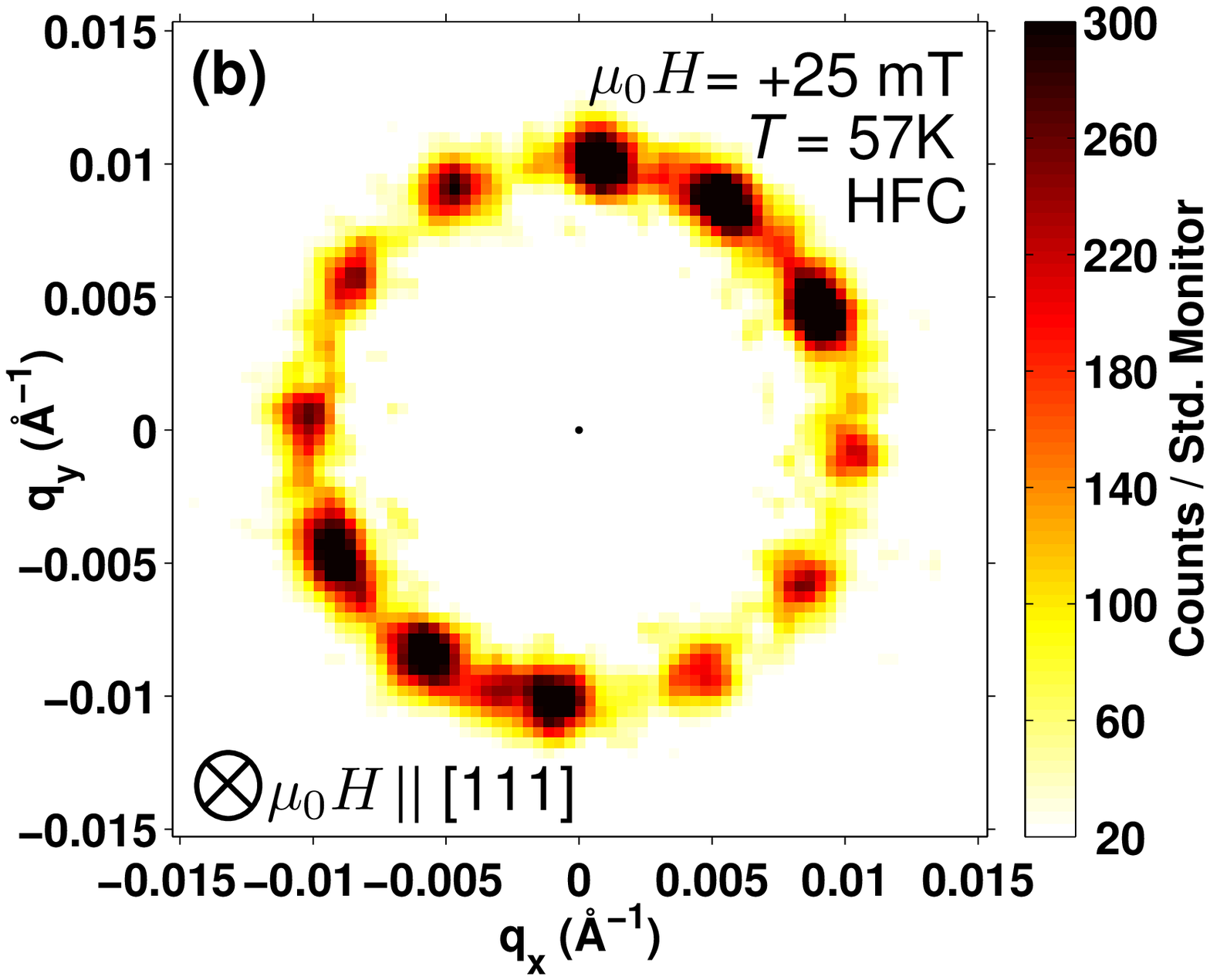}
  \caption{The SANS diffraction patterns collected at $T=57$~K and; (a) after cooling in zero applied fields and subsequently applying $\mu_{0}H \parallel$ [111] = +25~mT, (b) after magnetic field cooling in $\mu_{0}H \parallel$ [111] = +25~mT. No E-field is applied in these cases. The patterns are created by summing up all the measurements at different rotation angles in order to show all of the magnetic Bragg peaks in a single image. Since it was not possible to measure the rocking curve equivalently for all Bragg peaks, the diffraction spots may exhibit different intensities in the images. Statistical noise close to the origin of reciprocal space has been masked, and the data smoothed using a gaussian envelope that is smaller than the instrument resolution.}
  \label{fig:diff_patts_111}
\end{figure}

Fig.~\ref{fig:diff_patts_111}~(a) shows the SANS diffraction pattern collected in the SkL phase at $T=57$~K and $\mu_{0}H\parallel$ [111]~=~+25~mT. In this case the SkL was stabilised after initially cooling to $T=57$~K in zero applied fields, hereafter referred to as zero-field cooling (ZFC), and subsequently by applying +25~mT. The diffraction pattern is composed of six magnetic Bragg peaks distributed isotropically about the magnetic field axis that is characteristic of the triple-\textbf{q} SkL state~\cite{Muh09,Jon10,Ada12}. Consistent with previous studies, the equivalent magnetic wavevector for all six Bragg spots is determined to be \textbf{Q}$_{SkL}=0.0102(1)$~\AA$^{-1}$, thus giving a magnetic wavelength $\lambda_{SkL}=616(10)$~\AA.

In contrast to the ZFC approach, after instead magnetic field cooling (HFC) at +25~mT directly into the SkL phase, Fig.~\ref{fig:diff_patts_111}~(b) shows that we observe twelve magnetic Bragg peaks. Six of these peaks correspond to those peaks observed after ZFC (Fig.~\ref{fig:diff_patts_111}~(a)), and we conclude that after the HFC at least two SkL domains populate the sample. These data demonstrate that SkLs established inside our sample are sensitive to the magnetic field and temperature history; it is likely that the extra SkL domain observed in Fig.~\ref{fig:diff_patts_111}~(b) is metastable, and arises as a consequence of HFC directly into the SkL phase by crossing a narrow temperature window of critical spin fluctuations~\cite{Ada12}. A similar explanation by means of metastability was proposed in order to describe the observation of two coexisting SkL domains in Fe$_{1-x}$Co$_{x}$Si~\cite{Ada10}.

We also studied the application of E-fields in the $\mu_{0}H \parallel$ [111] geometry. For both the case of ZFC and HFC, applying sample voltages up to +444~V.mm$^{-1}$ \emph{within} the SkL phase gave no statistically observable effect on SkL, either in the spot positions or spot intensities. An alternative approach of cooling from 70~K to 57~K in simultaneous magnetic and E-fields (thus HFC and E-field cooling (EFC)), also showed no systematically reproducible effect on the SANS diffraction patterns, either in spot positions or intensities.

\begin{figure}
\centering
  \includegraphics[width=6.5cm]{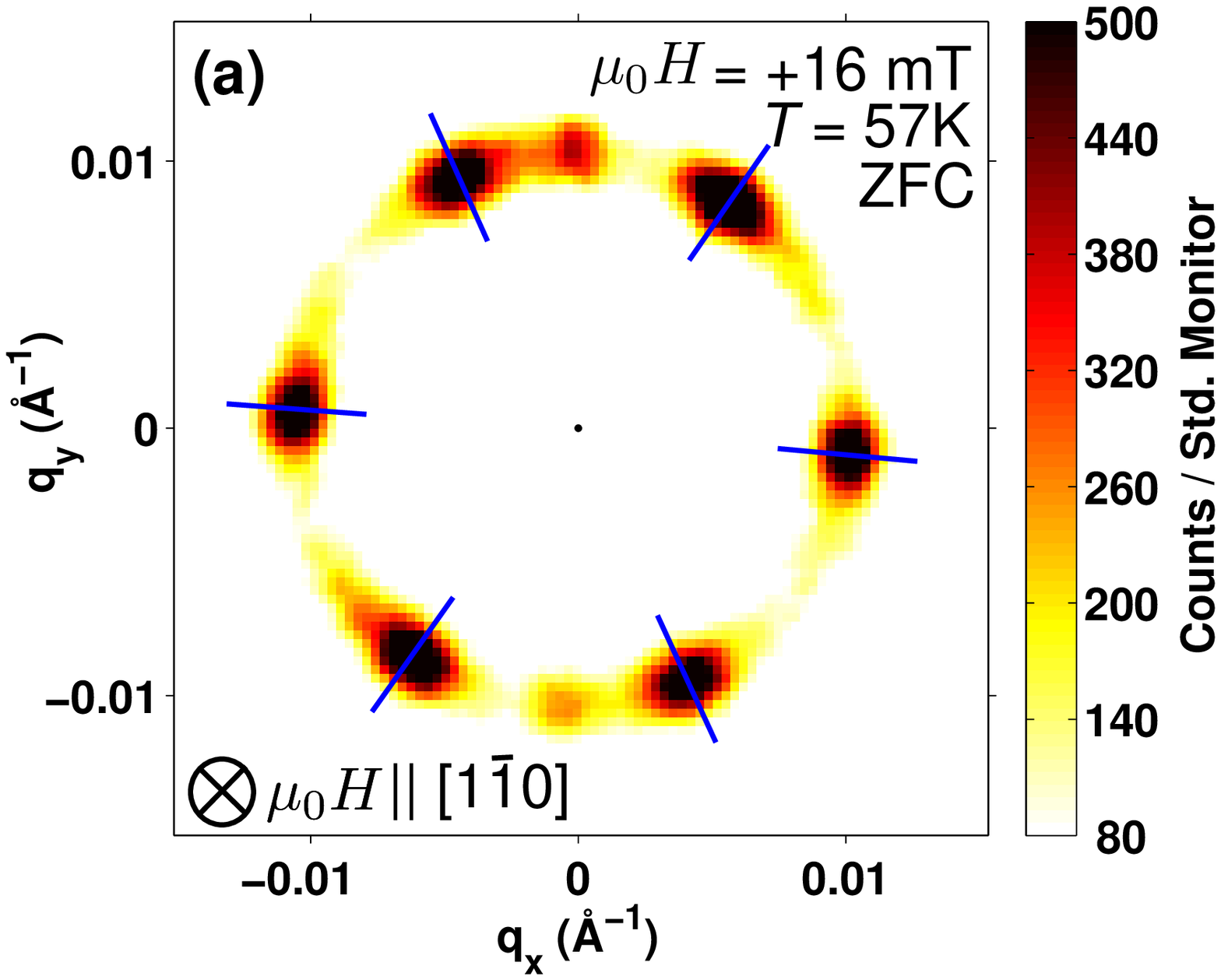}
  \includegraphics[width=6.5cm]{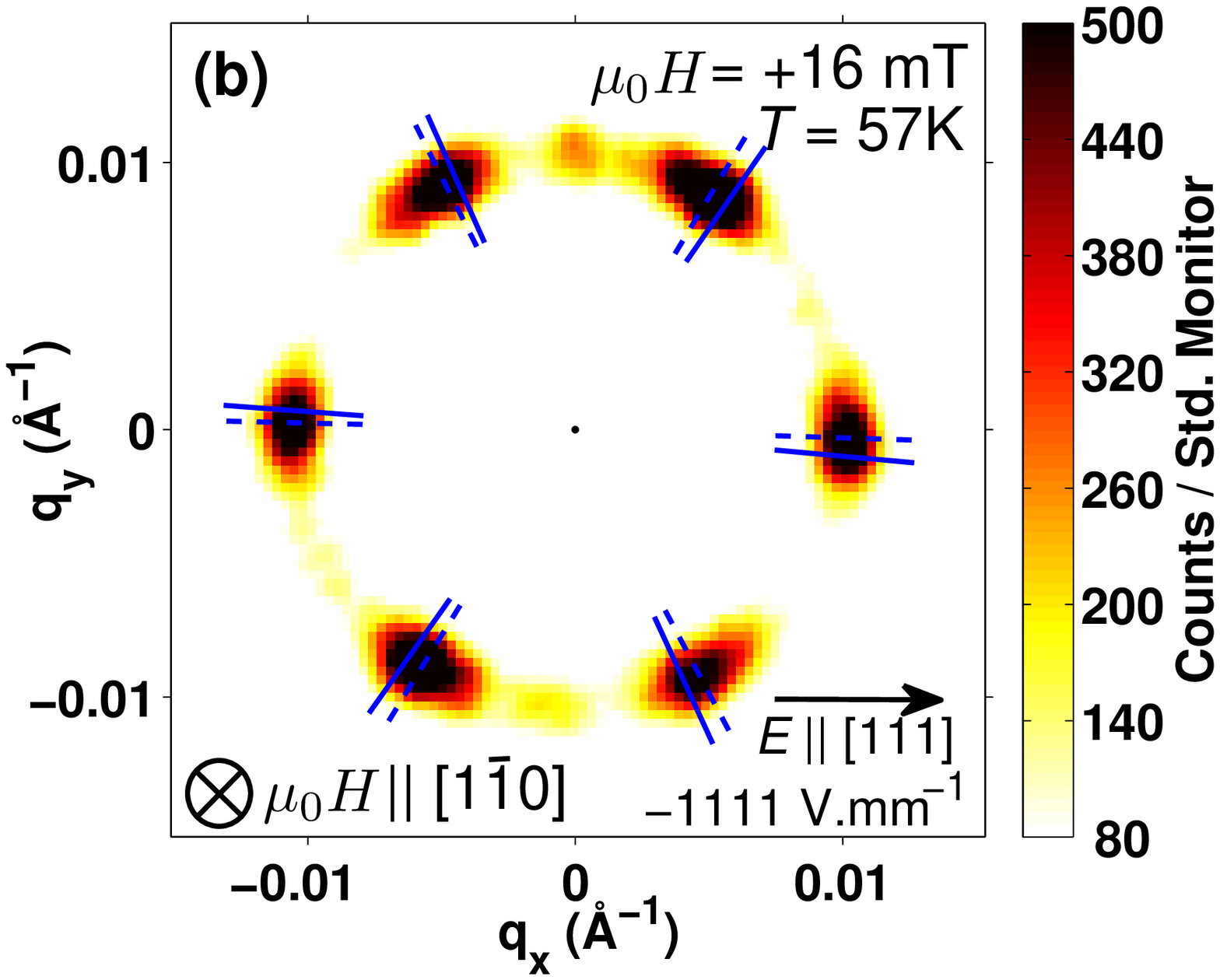}\\
  \vspace{6pt}
  \includegraphics[width=6.5cm]{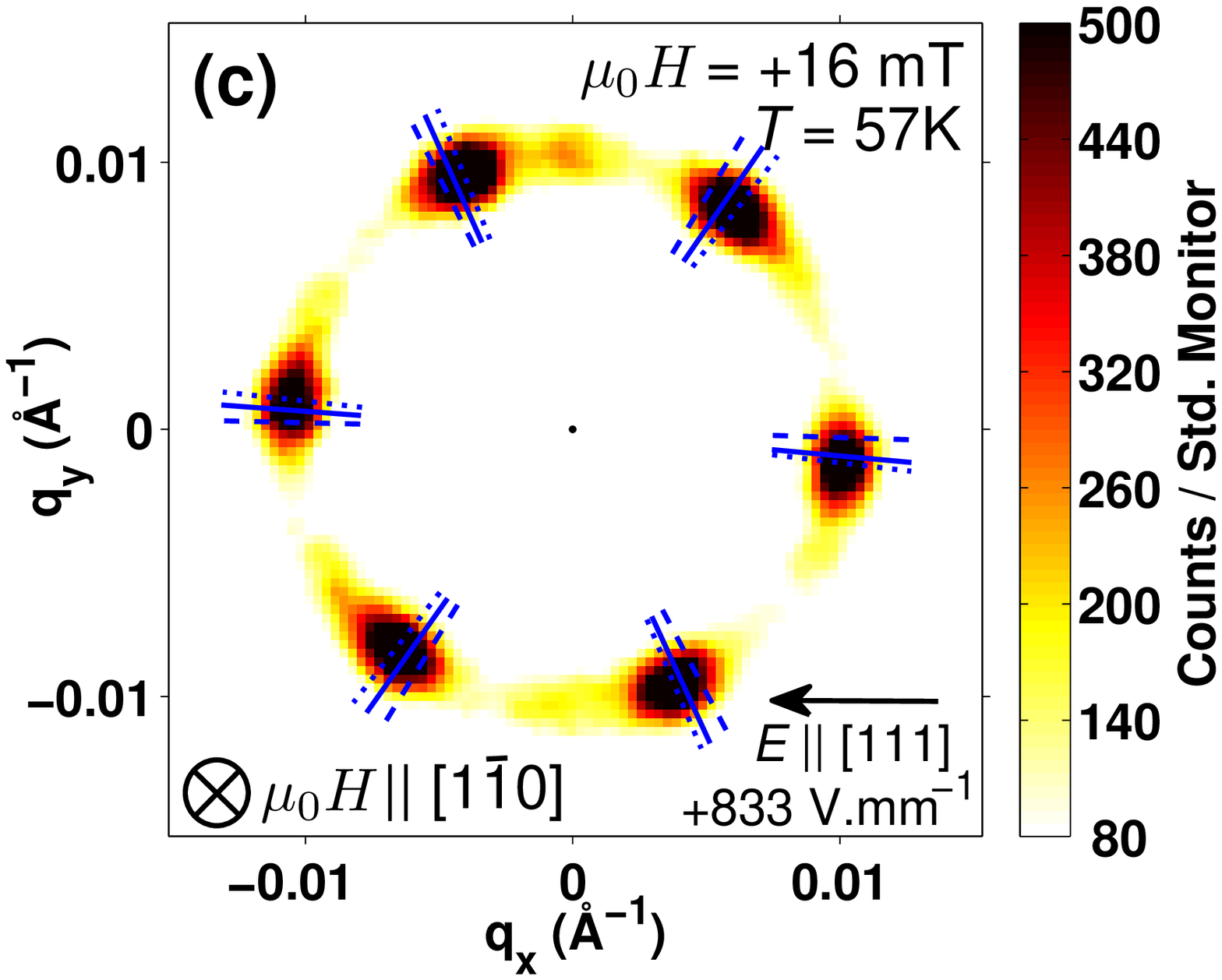}
  \includegraphics[width=6.5cm]{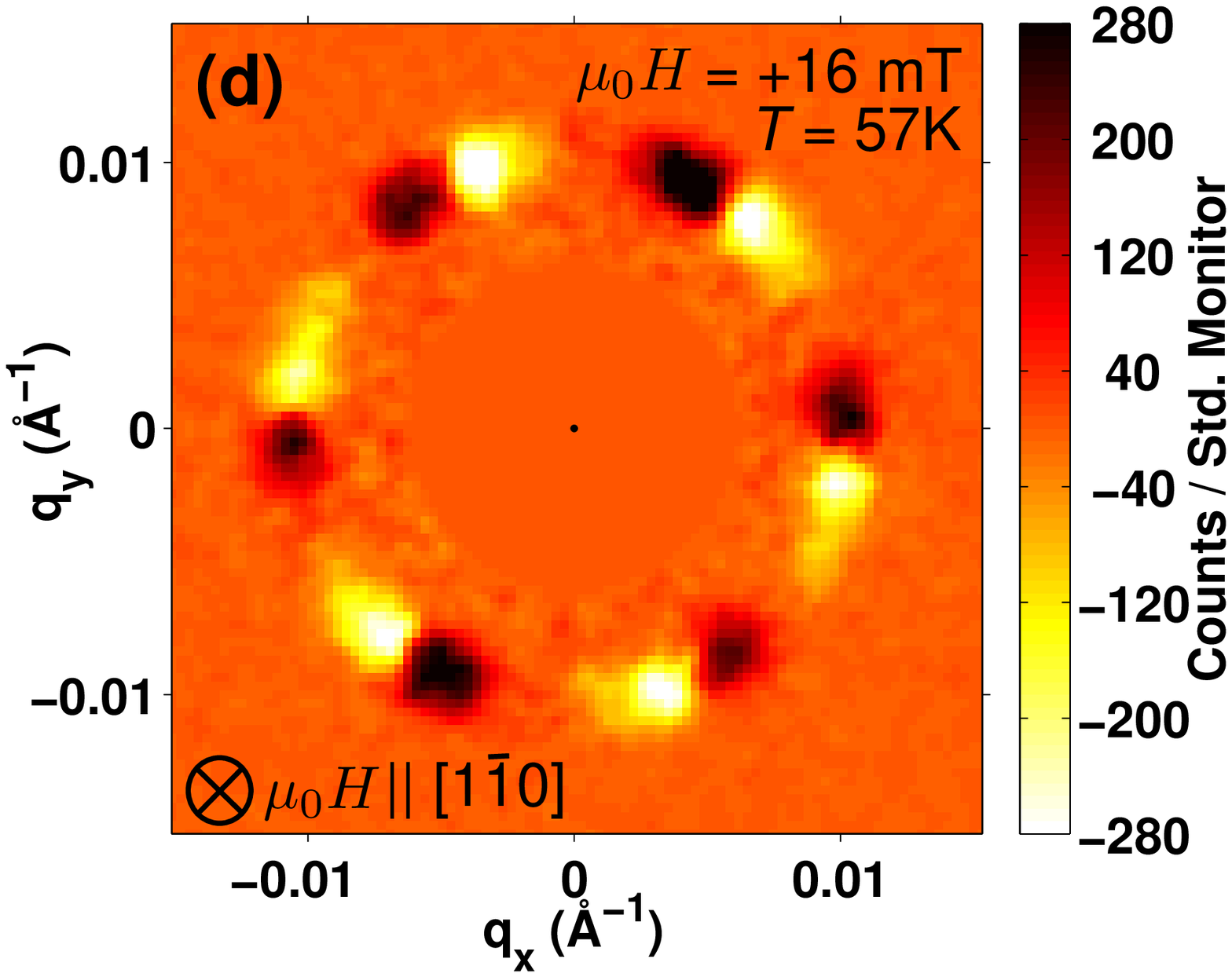}\\
  \vspace{6pt}
  \includegraphics[width=12cm]{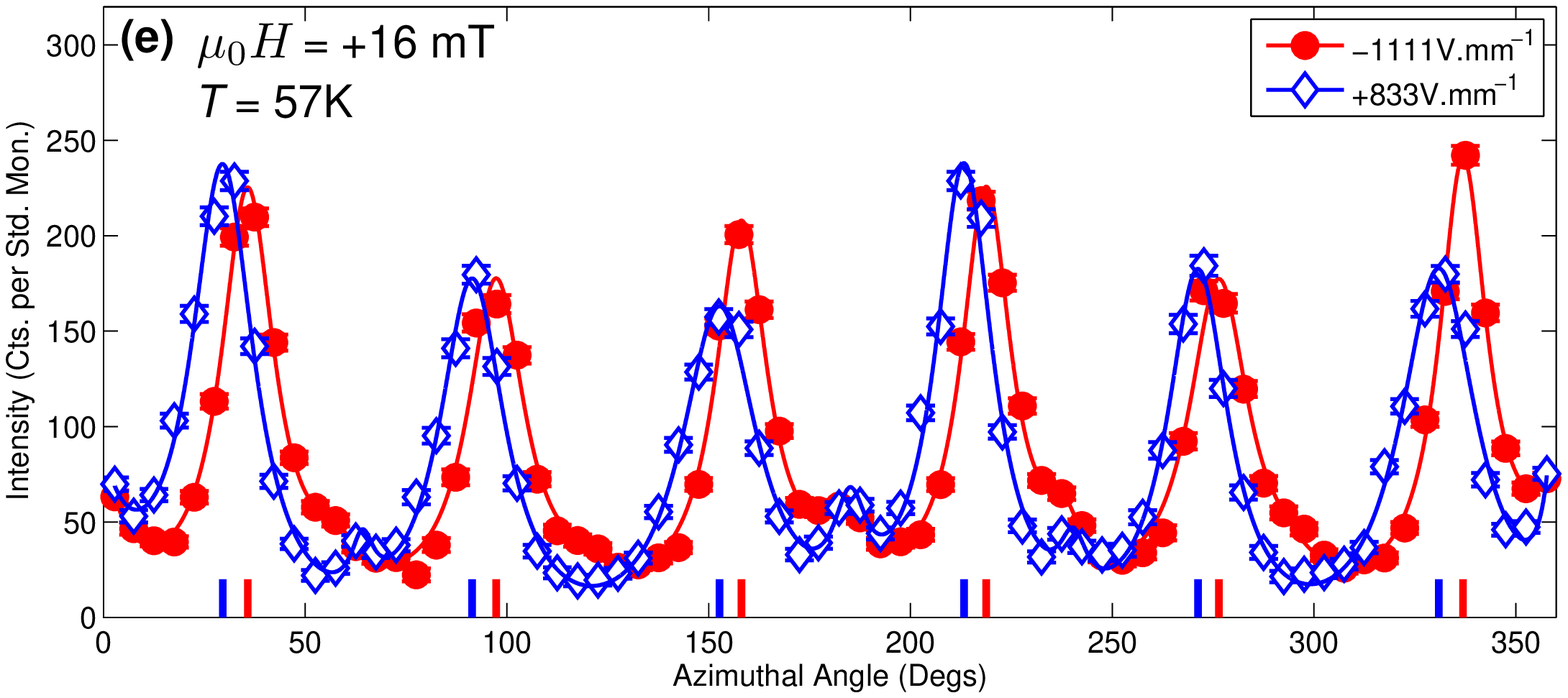}
  \caption{The SANS diffraction patterns collected at $T=57$~K and; (a) after ZFC and applying $\mu_{0}H \parallel$ [1\={1}0] = +16~mT, (b) after entering the SkL phase by applying $\mu_{0}H \parallel$ [1\={1}0] = +16~mT in an E-field of -1111~V.mm$^{-1}$ and, (c) after entering the SkL phase by applying +16~mT in an E-field of +833~V.mm$^{-1}$. In (a)-(c), the solid blue lines indicate the angular positions for the centres of mass for the diffraction spots of the pattern in (a). The centres of mass were determined by two-dimensional fits of each spot on the detector. In (b)-(c) the dashed lines show the angular positions of the Bragg spots for the pattern shown in (b). The dotted lines in (c) show the spot positions determined in an E-field of +833~V.mm$^{-1}$. In (d), we show a direct subtraction of the +833~V.mm$^{-1}$ foreground data from the -1111~V.mm$^{-1}$ foreground data. In (e) we present the azimuthal angle dependence of the diffracted intensity determined within a 2$\pi$ annulus sector that encompasses the diffraction spots in each of the patterns shown in (b) and (c). These data are binned every 5$^{\circ}$, with the 0$^{\circ}$ position corresponding to the vertical direction of the diffraction images. By fitting multiple-Lorentzian lineshapes to each dataset, the peak positions of the majority SkL domain in each case are determined, and indicated at the horizontal axis.}
  \label{fig:diff_patts_011}
\end{figure}

In the second geometry with $E\parallel$[111] and $\mu_{0}H \parallel$[1\={1}0], the magnetic field range over which the SkL is stabilised at 57~K was found to be lower than for the first geometry. We presume that this arises as a consequence of the lower apparent cross-section of the sample with respect to the applied magnetic field. This results in a demagnetisation effect so that a smaller magnetic field is sufficient to stabilise the SkL. Here, we do not attempt the numerical correction for the demagnetisation fields in our sample. Instead, we performed a magnetic field scan of the SkL phase in this geometry, and at $T=57$~K. We observed that in this geometry, and in our sample, i) the heart of the SkL phase was found for magnetic fields of magnitude close to 16~mT, and ii) the SkL orientation was not observed to transform as a function of magnetic field. Fig.~\ref{fig:diff_patts_011}~(a) shows the SkL stabilised in this second experimental geometry after ZFC to $T=57$~K, and subsequently applying a magnetic field of +16~mT~$\parallel$[1\={1}0]. As for the first geometry, we observe six strong diffraction spots due to the SkL, but of different orientation to that shown in Fig.~\ref{fig:diff_patts_111}~(a). The change in orientation of the SkL likely arises due a change in the effective magnetocrystalline anisotropy that tends to align the reciprocal SkL in Cu$_{2}$OSeO$_{3}$ with a $\langle100\rangle$ direction~\cite{Ada12}. Between the six strong peaks, some residual diffracted intensity due to a second minority SkL domain remains observable. We found that by HFC at +16~mT into the SkL phase, the relative population of the weaker domain increased, indicating a similar effect on crossing the regime of critical fluctuations as observed in the first geometry.

Next we discuss the effect of applied E-fields on the SkL in the $\mu_{0}H \parallel$[1\={1}0] geometry, which constitutes the main experimental finding of this paper. In what follows, it was found that after an initial ZFC to $T=57$~K, the SkL phase could be studied under different magnetic and electric field conditions by simply adjusting the magnetic field to cross the first-order SkL phase boundary, and without the requirement for cooling from the paramagnetic state. Each time after leaving the SkL phase, the cryomagnet was carefully degaussed to zero field, before the E-field condition was changed and the standard magnetic field of magnitude 16~mT re-applied to enter the SkL phase. This experimental approach was adopted in order to limit both the possible influences of sub-phases that may exist within the vicinity of the SkL phase~\cite{Wil11}, and also hysteresis effects; various control measurements showed the SANS diffraction patterns to be reproducible for SkLs prepared using this approach and under the same experimental conditions. Further comparison measurements after HFC (with or without a simultaneously applied E-field) showed that while the relative population of SkL domains differed compared to the ZFC approach, there was no difference between the two approaches of the observed E-field effect on the SkL domain shown in Fig.~\ref{fig:diff_patts_011}~(a), and that we now describe below.

Similarly as for the first geometry, applying dc E-fields of up to $+667$~V.mm$^{-1}$ \emph{within} the SkL phase produced no observable effect on the SANS diffraction pattern. In contrast however, Figs.~\ref{fig:diff_patts_011}~(b) and (c) show the diffraction patterns collected after increasing the magnetic field to +16~mT in the presence of \emph{poling} E-fields of $-1111$~V.mm$^{-1}$ and $+833$~V.mm$^{-1}$ respectively. We find that the effect of the E-field is to alter the precise orientation of the SkL, and that this is manifested in the diffraction patterns by an azimuthal rotation of the diffraction spots around the magnetic field axis in a direction dependent on the sign of the applied E-field. Fig.~\ref{fig:diff_patts_011}~(d) shows in a pictorial manner the relative rotation of the SkL rotation after subtracting the foreground data for panel (c) ($+833$~V.mm$^{-1}$) from the foreground data for panel (b) ($-1111$~V.mm$^{-1}$). A more quantitative comparison is shown in Fig.~\ref{fig:diff_patts_011}~(e) where we plot the azimuthal dependence of the diffracted intensity for the diffraction patterns shown in Figs.~\ref{fig:diff_patts_011}~(b) and (c). Here, the relative rotation of the \emph{entire} SkL is clearly evidenced in the data by a uniform shift in the Bragg peak positions between the two cases of SkLs created under positive and negative applied E-fields.

\begin{figure}
\centering
  \includegraphics[width=10cm]{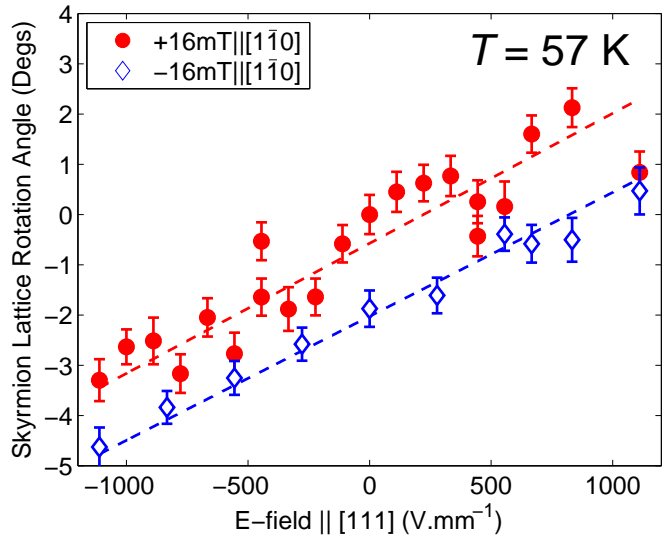}
  \caption{The E-field dependence of the relative rotation angle for the entire SkL, as determined by obtaining the mean azimuthal angular shift of all six peaks relative to the peak positions determined for $\mu_{0}H \parallel$[1\={1}0]~=~+16~mT and 0V. Each datapoint was collected after exiting the SkL phase via a degauss procedure, adjusting the E-field at zero magnetic field, and finally ramping the magnetic field to enter the SkL phase. The temperature always remained at $T=57$~K while these data were collected. Dashed lines show linear fits to the data for each polarity of the magnetic field.}
  \label{fig:dom_rotation}
\end{figure}

In Fig.~\ref{fig:dom_rotation} we present systematic measurements of the poling E-field-dependence of the SkL orientation relative to that displayed by the SkL in a magnetic field of +16~mT and zero E-field (Fig.~\ref{fig:diff_patts_011}~(a)). Data were collected under applied magnetic fields of magnitude 16~mT and of each polarity. We observe that the relative orientation of the entire SkL depends linearly on the the poling E-field in the range explored by our experiments ($\pm1111$~V.mm$^{-1}$). Surprisingly, and within uncertainty, we find identical linear fit gradients for each magnetic field polarity. For an applied magnetic field of +16~mT the linear fit gradient is 0.0023(3)~$^{\circ}$/(V.mm$^{-1}$), while for -16~mT we find a slope of 0.0022(1)~$^{\circ}$/(V.mm$^{-1}$).

\section{Discussion}


In order to clarify the origin of the E-field effect on the SkL that we observe in the $\mu_{0}H \parallel$[1\={1}0] geometry, we first mention that in the absence of external perturbations such as applied E-fields or electric currents, the SkL orientation is determined by a weak magneto-crystalline anisotropy~\cite{Muh09}. According to a recent SANS study of the SkL in Cu$_{2}$OSeO$_{3}$, with $\mu_{0}H\parallel$~[110] and no applied E-field~\cite{Sek12c}, the SkL orientation was observed to transform between single SkL domains related by a 30$^{\circ}$ rotation upon changing the applied magnetic field and temperature. This observation was suggested to arise due to a sensitive magnetic field and temperature-dependence of the underlying magneto-crystalline anisotropy. Since each of our measurements were done always at the same temperature and magnitude of the magnetic field, and the sign of the SkL rotation demonstrates a clear dependence on the sign of the E-field, we rule out a similar explanation here.

In the B20 alloy MnSi, a similar azimuthal rotation of the SkL diffraction pattern was demonstrated to be caused by the simultaneous application of both an electric current and a small temperature gradient within the SkL plane~\cite{Jon10}. The physical origin of the skyrmion motion was explained to arise due to forces exerted by the conduction electrons on the skyrmions~\cite{Jon10}. In order to overcome pinning and to initiate the skyrmion motion, it was found necessary to apply a current density of at least $j\sim2.2\times$10$^{6}$~A.m$^{-2}$. In our experiments on insulating Cu$_{2}$OSeO$_{3}$, in the presence of an applied voltage the residual current flow through the sample never exceeded a magnitude of $j\sim$~0.02~A.m$^{-2}$. Since this is eight orders of magnitude lower than that observed necessary to move the skyrmions in MnSi, we conclude that an alternative physical mechanism is necessary in order to explain our observations in Cu$_{2}$OSeO$_{3}$.



\begin{figure}
\centering
  \includegraphics[width=10cm]{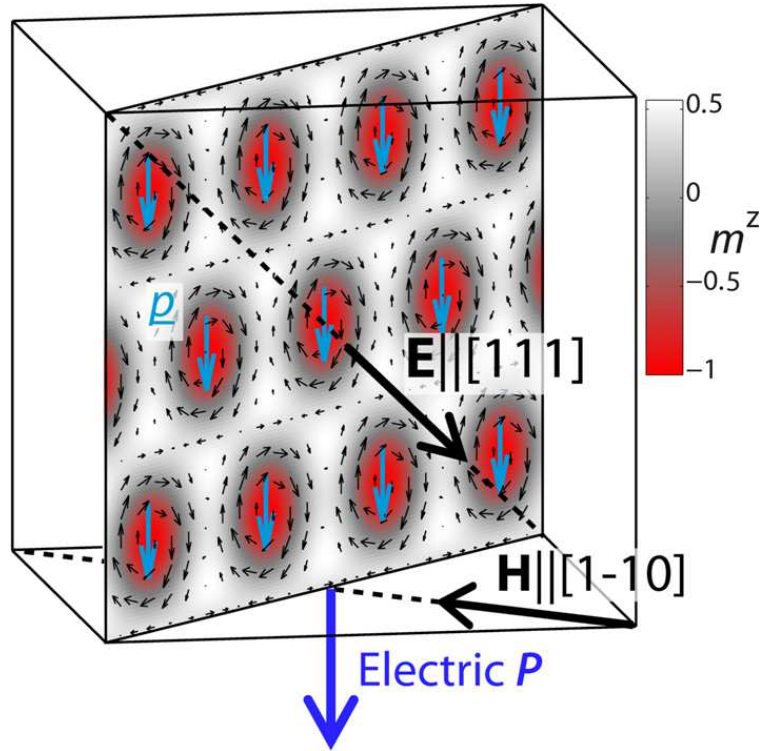}
  \caption{A schematic illustration showing the relative directions of the applied vector fields with respect to the cubic lattice of Cu$_{2}$OSeO$_{3}$. The SkL is shown to lie in the [1\={1}0] plane, and the E-field is applied along [111]. According to calculations for this direction of the applied magnetic field~\cite{Sek12b}, each skyrmion exhibits a local electric dipole $p$ along the [001] direction (light blue arrows). In zero E-field, the electric dipole is expected to lie 45$^{\circ}$ to [111].}
  \label{fig:physical_situation}
\end{figure}

At the heart of the observed E-field driven rotation of the SkL in Cu$_{2}$OSeO$_{3}$ is the physical origin of the ME interaction. Of the three microscopic mechanisms proposed to describe the magnetically-induced origin of electric polarisation in the magnetoelectrics and multiferroics, the more commonly observed exchange striction~\cite{Pic07} and spin-current/inverse DM mechanisms~\cite{Ser06,Kat05} are inactive in Cu$_{2}$OSeO$_{3}$. Instead, it has been shown in recent experimental~\cite{Sek12b} and theoretical~\cite{Sek12b,Yan12} work that the ME coupling in Cu$_{2}$OSeO$_{3}$ arises due to the less common spin-orbit (SO) mechanism~\cite{Jia06,Jia07,Ari07}. This mechanism is active between hybridised O $p$-ligands and Cu $d$-orbitals, leading to the emergence of a local electric dipole along the bond direction. Most intriguingly, and consistent with expectations based on symmetry arguments, the skyrmion spin texture permits a polar state only for certain directions of the applied magnetic field, namely $P\parallel$~[001] for $\mu_{0}H\parallel$~[110] and $P\parallel$~[111] for $\mu_{0}H\parallel$~[111]~\cite{Sek12,Sek12b}. For the case of $\mu_{0}H\parallel$~[110], a direct evaluation of the electric charge density expected due to the SO coupling mechanism shows that \emph{each} skyrmion supports a \emph{single local} electric dipole both within the SkL plane, and orthogonal to the applied magnetic field~\cite{Sek12b}. In Fig.~\ref{fig:physical_situation} we illustrate the situation expected to be realised for the case of our experimental geometry with $\mu_{0}H\parallel$~[1\={1}0]. In our experiments, each electric dipole is expected to make an angle~\cite{Sek12b} to the [111] direction along which the E-field is applied. Consequently, we propose that the intrinsic origin of the SkL rotation observed in our experiments (Fig.~\ref{fig:dom_rotation}) is due to a direct coupling between the E-field and the local electric dipole at the heart of each skyrmion.

As is evident from Fig.~\ref{fig:dom_rotation}, and over the explored range, the terms that cause the shift in orientational free energy of the SkL scale linearly with the E-field. Furthermore, we find that there is no `critical' E-field below which the SkL orientation remains determined by the magneto-crystalline anisotropy, and above which the SkL starts to rotate. This latter observation lies in contrast to that seen on MnSi, whereby the SkL was only observed to rotate for current densities above a critical threshold~\cite{Jon10}. It may seem natural to discuss the SkL rotation in Cu$_{2}$OSeO$_{3}$ as arising due to a torque generated by the E-field on each local electric dipole, yet our experiments revealed no effect on the SkL by changing the E-field \emph{within} the SkL phase. While future experiments will explore the possible manipulation of the skyrmions within the SkL phase, from the present study we conclude that the E-field does not rotate the SkL in a dynamic way as is achieved by changing the applied current density in MnSi.~\cite{Jon10} Instead, by entering the SkL phase in the presence of an E-field, we expect that the SkL nucleates in an orientation that minimises the free energy with respect to both the magneto-crystalline anisotropy, and the electrostatic potential experienced by the local electric dipoles. This proposal requires the dipole and skyrmion lattice orientations to be both coupled, and to display preferred orientations with respect to one another. These properties may be expected to arise due to the chiral interactions evident in the host system, and our proposal invites a detailed theoretical analysis.





The emergence of a local electric dipole at the centre of each skyrmion ascribes to each a particle-like property which may allow the manipulation of individual skyrmions~\cite{Yu10}. Since each skyrmion is expected to display an electric dipole for $\mu_{0}H\parallel$~[1\={1}0], the observed E-field effect may be expected even though the E-field was $\parallel$~[111], and so not perfectly $\parallel$~[001] where the polarisation is observed experimentally~\cite{Sek12b}. We do, however, only expect an influence of the poling E-field for cases where the magnetic field is applied along a crystal direction that permits the emergence of the local electric dipoles. This suggestion needs to be validated experimentally, yet the picture proposed here is consistent with the apparent absence of a clear SkL rotation in our first geometry with $E\parallel\mu_{0}H\parallel$~[111]. Since in this situation, the orientation of the local electric dipole is calculated to be parallel to [111]~\cite{Sek12b}, its energy is minimised with respect to the direction of the E-field, and the SkL orientation can not be influenced.


Finally, we clarify that the independence of the sign of SkL rotation on the magnetic field polarity (Fig.~\ref{fig:dom_rotation}) is consistent with results reported in other work~\cite{Sek12b,Yan12}. As shown in measurements of the magnetic-field-induced electric polarisation~\cite{Sek12,Sek12b}, for both $\mu_{0}H\parallel$~[111] and $\mu_{0}H \parallel$[110], the electric polarisation that emerges varies with the square of the magnetic field, so that it displays the same sign and magnitude as a function of magnetic field for both polarities. This behaviour was shown to confirm the SO coupling model as the microscopic origin of the ME coupling~\cite{Sek12b}, and so indicates that for $\mu_{0}H \parallel$[110] the sign of the local electric dipole carried by each skyrmion is independent from the magnetic field polarity.

\section{Conclusion}
In conclusion, using small-angle neutron scattering and simultaneously applied magnetic and electric fields, we have demonstrated the electric field control of the skyrmion lattice (SkL) in Cu$_{2}$OSeO$_{3}$. With $\mu_{0}H\parallel$~[1\={1}0], and $E\parallel$~[111], by poling into the SkL phase we show how the precise orientation of the SkL may be controlled in a manner dependent on both the size and sign of the applied E-field. This observation reveals the inherent magnetoelectric coupling between the applied E-field and the individual skyrmions, and demonstrates how the generally emergent properties of skyrmions will be dependent on the properties of the host system.


\section{Acknowledgements}
Technical support from M.~Zolliker is gratefully acknowledged. Experiments were performed at the Swiss spallation neutron source SINQ,
Paul Scherrer Institute, Villigen, Switzerland, and are supported by the Swiss National Science Foundation and its NCCR MaNEP, the CONQUEST (Controlled Quantum Effects and Spin Technology) grant of the European Research Council (ERC), and the Croatian Science Foundation contract No. 02.05/33.

\section{References}
\bibliographystyle{unsrt}
\bibliography{cu2oseo3_sans}
\end{document}